# Relaxations as key to the magnetocapacitive effects in the perovskite manganites

F. Schrettle,[1] P. Lunkenheimer,[1,*] J. Hemberger,[1] V. Yu. Ivanov,[2] A. A. Mukhin,[2]
A. M. Balbashov,[3] and A. Loidl[1]

[1]*Experimental Physics V, Center for Electronic Correlations and Magnetism, University of Augsburg, D-86135 Augsburg, Germany*
[2]*General Physics Institute of the Russian Academy of Sciences, 38 Vavilov Street, 119991 Moscow, Russia*
[3]*Moscow Power Engineering Institute, 14 Krasnokasarmennaja Street, 111250 Moscow, Russia*

We present a detailed dielectric study of the relaxation effects that occur in several perovskite rare-earth manganites, including the multiferroics $TbMnO_3$ and $DyMnO_3$. We demonstrate that the strong magnetocapacitive effects, observed for electrical fields $\mathbf{E}||\mathbf{c}$, are nearly completely governed by magnetic-state induced changes of the relaxation parameters. The multiferroic materials, which undergo a transition into a spiral magnetic state, show qualitatively different relaxation behavior than those compounds transferring into an A-type antiferromagnetic state. We ascribe the relaxations in both cases to the off-center motion of the manganese ions, which in the multiferroic systems also leads to the ferroelectric ordering.

In the recent revival of the magnetoelectric effect [1], perovskite rare-earth manganites have played a prominent role [2,3]. For some of these materials like $TbMnO_3$, ferroelectric (FE) polarization is induced by the onset of non-collinear helical spin order which breaks inversion symmetry [2,3,4,5]. Two mechanisms have been proposed to explain the occurrence of ferroelectricity: A purely electronic one invoking spin currents, which establish FE polarization from the electronic charge density only [6] and a second one, where the oxygen ions are shifted off the centrosymmetric positions [7]. However, in recent first-principle studies of $TbMnO_3$ [8] it has been shown that i) only the ionic-displacement model can explain the size of the observed FE polarization and that ii) $Mn^{3+}$ and $Tb^{3+}$ displacements are generally greater than those of the oxygen ions.

In several seminal publications [2,3,9,10], the phase diagram and magnetocapacitive behavior of the perovskite rare-earth manganites have been investigated in detail. Spectacular effects are found especially for $RMnO_3$ with $R$ = Eu, Gd, Tb, and Dy [2,3,10,11]. Below about $T_N$ = 40-50 K, the Mn 3d spins in these materials undergo a collinear sinusoidal antiferromagnetic (AF) ordering. At lower temperatures, either an A-type AF (for $R$ = Eu and Gd) or a transverse-spiral spin state accompanied by FE polar order (for $R$ = Tb and Dy) is assumed [2,3,10]. The Eu compound does not become FE. For $GdMnO_3$, which is close to the phase boundary between the two ground states, contradictory results were reported [10,12] but at least it is clear that FE ordering can be induced by a weak external field $\mathbf{H}||\mathbf{b}$. All these compounds exhibit significant magnetocapacitive effects. The strong variation of the magnetic characteristics with $R$-ion arises from an increase of tilting between the $MnO_6$ octahedra with decreasing ion sizes influencing the competing magnetic interactions. It can be also varied by doping: Recently it was shown that doping $EuMnO_3$ with Y drives the system from A-type AF to spiral spin order [11].

In the perovskite manganites, one of the most prominent magnetocapacitive effects is a strong anomaly of the temperature-dependent dielectric constant $\varepsilon'$ at the transition into the spiral or A-type AF state. Here either a steplike behavior or sharp peaks are observed, the latter having been taken as evidence for FE ordering. These effects arise along the crystallographic $\mathbf{a}$ and/or $\mathbf{c}$ direction, but not along $\mathbf{b}$. This may be understood by considering the symmetry predictions made within the proposed models [6,7], namely that the polarization $\mathbf{P} \propto \mathbf{e} \times \mathbf{Q}$ with the spiral axis $\mathbf{e}$ and the propagation vector of spin order $\mathbf{Q}$. Interestingly, in all compounds for electrical field direction $\mathbf{E}||\mathbf{c}$ a well-pronounced relaxation feature is observed to be superimposed to the mentioned anomalies in $\varepsilon'(T)$ [3,11]. It shows up as a steplike increase of $\varepsilon'$, shifting to higher temperatures with increasing frequency. So far, this feature was widely neglected and in many cases $\varepsilon'$ was shown for a single frequency only. The origin of this relaxation is unclear so far. In the present work, we provide a detailed investigation of this $\mathbf{c}$-axis relaxation in various rare-earth perovskite manganites ($R$ = Eu, Gd, Tb, Dy) and the doped system $Eu_{1-x}Y_xMnO_3$. We show that the relaxation effects play an important role for the magnetocapacitive behavior and can give some clues on the nature of the FE ordering in the multiferroic manganites.

Single crystals were prepared as described in [11,13] and silver-paint contacts were applied to achieve a field direction $\mathbf{E}||\mathbf{c}$. The dielectric properties were determined using a frequency-response analyzer and an LCR meter [14]. For measurements between 1.8 K and 300 K and in external magnetic fields up to 5 T a Quantum-Design Physical-Property-Measurement-System was used.

Figure 1 shows $\varepsilon'(T)$ at various frequencies in $DyMnO_3$ and $GdMnO_3$, the latter in a magnetic field $\mathbf{H}||\mathbf{c}$ of 1 T. In both cases, the behavior is dominated by a typical relaxation feature: $\varepsilon'$ decreases from a high plateau value,

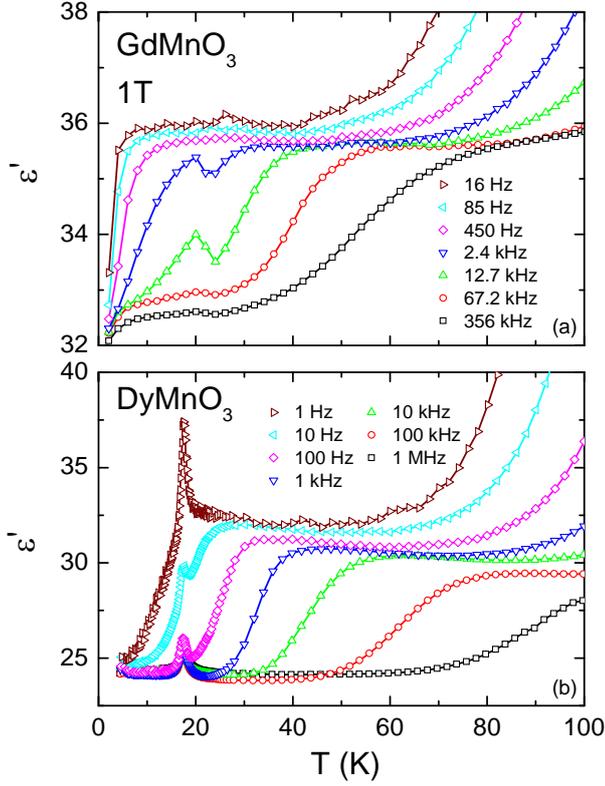

FIG. 1 (color online). $\varepsilon'(T)$ of GdMnO$_3$ at a magnetic field **H**||**c** of 1 T (a) and DyMnO$_3$ (b) for various frequencies. The lines are guides to the eyes. The absolute values of $\varepsilon'$ have relatively high uncertainty due to an ill-defined geometry.

$\varepsilon_s$, to a lower one, $\varepsilon_\infty$, because with decreasing temperature the relaxing entities can no longer follow the applied ac field [15]. At the mid-point of the step, the condition $\nu = 1/(2\pi\tau)$ is fulfilled with $\tau$ the relaxation time [15]. Its temperature dependence explains the observed frequency shift of the $\varepsilon'$-step. The additional increase of $\varepsilon'(T)$ at high temperatures is of extrinsic origin and caused by electrode polarization [16]. The relaxation steps observed by us quantitatively agree with literature results [3], which were obtained on crystals prepared by different crystal growers. In addition, these relaxations are found for **E**||**c** only. Thus, a non-intrinsic origin seems very unlikely [17]. In agreement with Refs. [3,10], in both materials there are distinct anomalies in $\varepsilon'(T)$, superimposed to the relaxation steps. They occur at the transition into the spiral magnetic (DyMnO$_3$, $T_c \approx 17.5$ K) or A-type AF (GdMnO$_3$, $T_c \approx 22$ K) state [18]. In contrast, the paramagnetic/sinusoidal-AF phase transition ($T_N \approx 41$ K [19] and 39 K [3], respectively) seems to have no influence on $\varepsilon'$. At intermediate frequencies, for both systems a peak shows up in $\varepsilon'(T)$. However, while at low and high frequencies in GdMnO$_3$ [Fig. 1(a)] almost no anomaly is found, in DyMnO$_3$ [Fig. 1(b)] the peaks become especially well pronounced. In Refs. [2,3] such peaks were taken as indication of a FE transition in TbMnO$_3$ and DyMnO$_3$. For the Gd compound, neither the upper nor the lower $\varepsilon'(T)$ plateau show any anomaly and thus $\varepsilon_s$ and $\varepsilon_\infty$ are constant. In contrast, in DyMnO$_3$ $\varepsilon_s$ and $\varepsilon_\infty$ are sensitive to the transition. The much stronger peak at 1 Hz implies that the relaxation strength $\Delta\varepsilon = \varepsilon_s - \varepsilon_\infty$ varies at $T_c$.

A qualitatively similar behavior as in Fig. 1(a), was also found in GdMnO$_3$ for higher magnetic fields up to 5 T and in Eu$_{1-x}$Y$_x$MnO$_3$ with $x = 0$ and 0.1. In the further course of this work, these systems will be called "type A". A behavior as shown in Fig. 1(b) was also revealed in TbMnO$_3$ and in Eu$_{1-x}$Y$_x$MnO$_3$ for $x = 0.2$, 0.3, and 0.5, which we term "type B". Obviously, all systems of type B exhibit a transition into a FE spiral spin state while those of type A transfer into an A-type AF state.

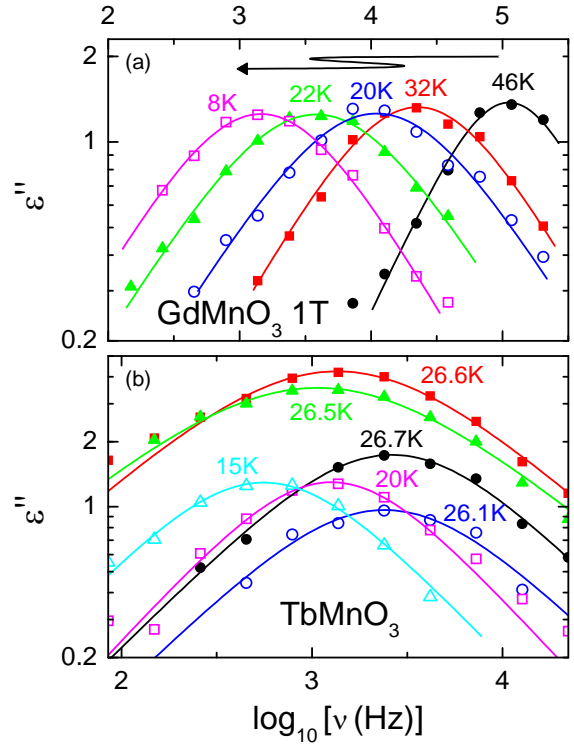

FIG. 2 (color online). $\varepsilon''(\nu)$ of GdMnO$_3$ at a magnetic field **H**||**c** of 1 T (a) and of TbMnO$_3$ (b) for selected temperatures. The lines are fits with the Cole-Cole function. The arrow in (a) indicates the shift of the peak frequency.

Since $\varepsilon_s$ and $\varepsilon_\infty$ of GdMnO$_3$ remain constant at $T_c$, the only way to explain the observed peaks at intermediate frequencies is by assuming a jump of the relaxation time. This should lead to a horizontal shift of the $\varepsilon'(T)$-steps, consistent with the data of Fig. 1(a). Relaxation times can be best determined in plots of the frequency-dependent dielectric loss, $\varepsilon''$, which should show a peak when $\nu = 1/(2\pi\tau)$ [15]. In Fig. 2(a), such plots are provided for GdMnO$_3$ for temperatures around the transition. Between the highest temperatures and 22 K, the loss peaks shift to lower frequencies, mirroring the expected slowing down of the relaxational dynamics. However, at 20 K, just below $T_c$, the loss peak is located at a higher frequency than for 22 K. Thus, the relaxation accelerates at the phase transition. For further decreasing temperature, it again resumes its conventional slowing down. All other investigated type A systems show similar behavior. Also in the type B systems, $\tau$ varies at the phase transition. As an example, in Fig. 2(b) we show $\varepsilon''(\nu)$ for TbMnO$_3$. Again, around $T_c \approx 26.5$ K, an unconventional $\tau(T)$ behavior shows up, analyzed in detail below. In addition, in contrast to the type A systems, also the peak height significantly changes at $T_c$. This again demonstrates the above-mentioned variation of $\Delta\varepsilon$, which is correlated with the peak amplitudes [15].

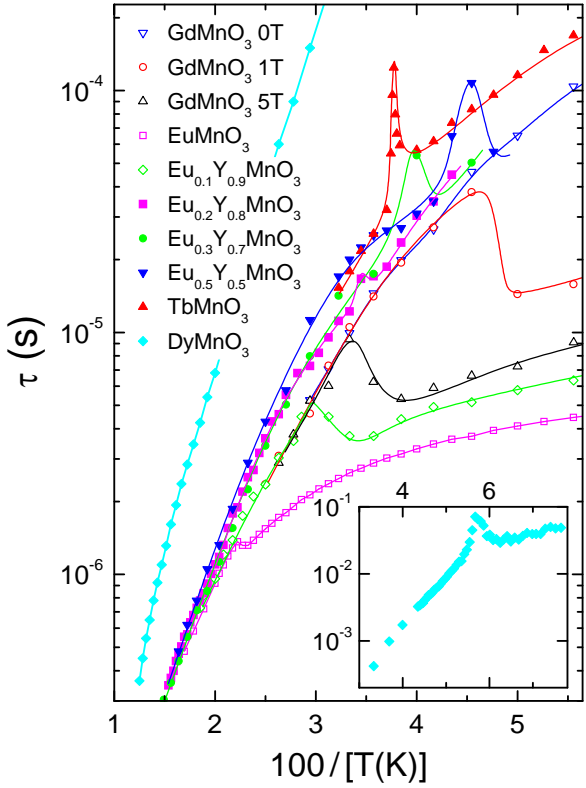

FIG. 3 (color online). Arrhenius plot of $\tau$ of all investigated compounds. The lines are guides to the eyes. The inset shows the results for DyMnO$_3$, beyond the scale of the main frame.

In all investigated materials, the loss peaks are broader than for the so-called Debye case where a single $\tau$ is assumed for all relaxing entities [15]. Such a broadening usually is ascribed to a distribution of relaxation times and well known, e.g., for disordered systems [15]. The curves in Fig. 2 can be well fitted by the commonly-employed empirical Cole-Cole function, $\varepsilon'' = \mathrm{Im}\,[\Delta\varepsilon/(1+(i2\pi\nu\tau)^\alpha)]$ [15]. Here $\alpha$ is a width parameter; $\alpha = 1$ implies the Debye case. Values of $\alpha < 1$ lead to a symmetric broadening of the loss peaks with power laws of $\nu^{1-\alpha}$ and $\nu^{\alpha-1}$ at the low- and high-frequency flanks of the peaks. The resulting relaxation times are shown in Fig. 3 in an Arrhenius representation, together with those of the other materials. There is a marked qualitative difference concerning the behavior at $T_c$ of type A and B systems: Coming from high temperatures, the first ones (open symbols) show a steplike decrease of $\tau$, i.e. at $T_c$ the relaxation becomes *faster*. In contrast, the type B systems (closed symbols) exhibit a sharp peak, i.e. at $T_c$ the relaxation first becomes *slower* and then again accelerates to resume the previous behavior. While a slowing down at the transition is often found in FE materials [20], a steplike behavior is uncommon for dielectric relaxations in any materials. Below and above $T_c$, $\tau(T)$ does not follow simple thermally activated behavior, which would result in a linear increase in Fig. 3. A bending of $\tau(1/T)$ curves as in Fig. 3 sometimes is taken as indication of tunneling processes at low temperatures. The fact that almost all curves join at high temperatures (for DyMnO$_3$ this may occur at higher temperatures only) indicates that the relaxation has the same origin in all systems. Another point to be noted is the influence of the magnetic field in GdMnO$_3$: Above the transition, $\tau$ remains unaffected by the field, but below $T_c$ the relaxation times strongly depend on $H$. For $H = 0$ T, no anomaly is seen but for 1 and 5 T the mentioned step in $\tau$ occurs, mirroring the well-known shift of the A-type AF transition with magnetic field [10].

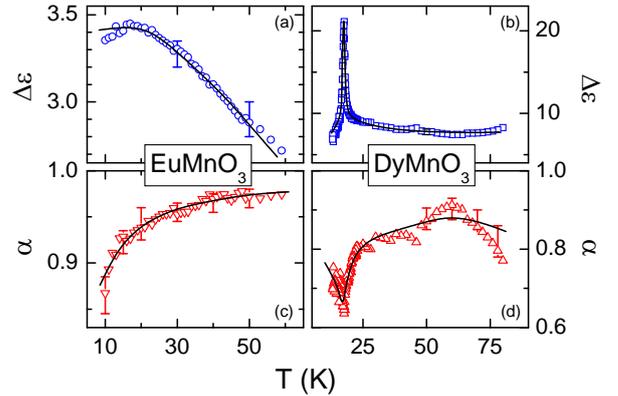

FIG. 4 (color online). Temperature dependence of $\Delta\varepsilon$ (a, b) and $\alpha$ (c, d) of EuMnO$_3$ and DyMnO$_3$. The lines are guides to the eyes.

In Fig. 4, we provide the relaxation strength and width parameters, again for a type A [(a) and (c)] and a type B system [(b) and (d)]. For type A systems, both quantities show no anomaly at $T_c$ ($\approx 43$ K for the example of EuMnO$_3$). The relaxation becomes stronger for lower temperatures but it does not follow the frequently-found

Curie law $\Delta\varepsilon \sim 1/T$. Relaxation widths often approach the Debye case ($\alpha = 1$) for high temperatures, which also is found in the present case. The type B systems show a strong peak in $\Delta\varepsilon$ at $T_c$ [e.g., Fig. 4(b)]. Their width parameter exhibits a minimum at $T_c$ [e.g., Fig. 4(d)], i.e. the peaks become more broadened, possibly due to fluctuation-induced disorder at the phase transition leading to a broader distribution of relaxation times.

Overall, our results clearly reveal that the relaxation plays a dominant role for the magnetocapacitive behavior of the rare-earth manganites. The only parameter influenced by the transition at $T_c$ that is not directly related to the relaxation is $\varepsilon_\infty$ in the type B systems (for type A, it is constant). It can be deduced, e.g., from Fig. 1(b) at $\nu \geq 1$ kHz and varies by about 6%, only. In contrast, the relaxation parameters $\tau$ and $\Delta\varepsilon$ vary up to a factor of two! Also for type A systems, the $\tau$ variation is of similar magnitude. These anomalies lead to relatively moderate changes of $\varepsilon'$ at $T_c$ only (e.g., Fig. 1) but in fact something dramatic is happening here. Also the strong variation of $\varepsilon'$ with magnetic field in these materials [3,10] can be traced back to a change of the relaxation parameters. For example, in Fig. 3 it is demonstrated that in GdMnO$_3$ at low temperatures, a field of 5T accelerates the relaxation by more than a factor of 10.

We summarize the experimental observations of this work: i) A **c**-axis relaxation is observed in all investigated manganites, ii) the relaxation significantly accelerates in non-FE type A systems when passing into the canted AF state and undergoes a critical slowing down in the type B systems when passing into the FE helical spin state, and iii) the relaxation strength in type A systems has a smooth temperature dependence, while in type B systems it becomes critically enhanced at the FE transition. Concerning the microscopic origin of the observed relaxation, it should be noted that in canonical ferroelectrics of order-disorder type, the occurrence of a relaxation is well known [20]. This relaxation shows a critical slowing down at $T_c$, quite similar to the behavior in the type B systems and is caused by dipole-like entities within a double-well potential. If the dipolar entities are coupled the system can undergo an order-disorder FE phase transition. If the coupling is weak the dipoles freeze in, devoid of long-range order. We propose that all rare-earth manganites exhibit a shallow double-well potential of the manganese ions along **c**. Based on the dipolar strength of the relaxation as documented in Figs. 1, 2, and 4, we estimate that the separation of the potential minima must be well below 0.01 Å. We conclude that all perovskite manganites are close to a FE order-disorder instability. However, the multiferroic manganites need magnetic support via spiral spin order to finally establish long range polar order. In all other systems this dipolar relaxation is frozen in. This idea is experimentally supported by Figs. 3 and 4(b) documenting that close to the FE phase transition the relaxation strength is critically enhanced and the relaxation time slows down, as usually observed in order-disorder ferroelectrics.

In summary, we found that the rare earth manganites are governed by so far widely neglected relaxational effects, most probably signaling the motion of manganese ions in a shallow double-well potential along **c** within the oxygen octahedra. In the non-FE type A systems, the relaxation time is strongly decreased at the onset of the canted AF phase but finally the dipolar reorientation freezes in on further cooling. In systems where spiral spin order is established this order-disorder phenomenon is strongly enhanced resulting in long-range FE order.

Finally we want to remark that the dielectric relaxation, very recently detected at the spin-flop transition in DyMnO$_3$ for **E**||**a** and **H**||**b** [21], also displays all typical features of the critically slowing down relaxations observed in canonical order-disorder ferroelectrics.

This work was supported by the Deutsche Forschungsgemeinschaft via the Sonderforschungsbereich 484 and the Russian Foundation for Basic Researches (project 06-02-17514).

———————